\documentstyle[epsfig]{aipproc}

\begin{document}

\title{W-Pair Production with YFSWW/KoralW%
      \thanks{Work partly supported by the Maria Sk\l{}odowska-Curie
              Joint Fund II PAA/DOE-97-316 and
              by the US Department of Energy Contracts DE-FG05-91ER40627
              and DE-AC03-76ER00515.}
}

\author{W. P\l{}aczek$^{a,b}$, S. Jadach$^{c,d,b}$, M. Skrzypek$^{d,b}$ 
        B.F.L. Ward$^{e,f,b}$  \\ and Z. W\c{a}s$^{d,b}$}
\address{
$^a$  Institute of Computer Science, Jagellonian University,\\
        ul. Nawojki 11, 30-072 Cracow, Poland,\\
$^b$CERN, CH-1211 Geneva 23, Switzerland,\\
$^c$DESY Zeuthen, Platanenallee 6, D-15738 Zeuthen, Germany\\
$^d$Institute of Nuclear Physics,
  ul. Kawiory 26a, 30-055 Cracow, Poland,\\
$^e$Department of Physics and Astronomy,\\
  The University of Tennessee, Knoxville, TN 37996-1200, USA,\\
$^f$SLAC, Stanford University, Stanford, CA 94309, USA.
}

\maketitle

\vspace{-2mm}
\begin{abstract}
A theoretical description of W-pair production in terms of two complementary
Monte Carlo event generators {\tt YFSWW} and {\tt KoralW} is presented. 
The way to combine the results of these two programs in order to get 
precise predictions for WW physics at LEP2 and LC energies is discussed. 
\end{abstract}

\vspace{-2mm}

The process of W-pair production in electron--positron colliders is very
important for testing the Standard Model (SM) and searching for signals of
possible ``new physics''; see e.g.~Ref~\cite{LEP2YR:1996}. 
One of the main goals of investigating this process at present and future
$e^+e^-$ experiments is to measure precisely the basic properties of the
W boson, such as its mass $M_W$ and width $\Gamma_W$. This process also 
allows a study, at the tree level, of triple and quartic gauge boson couplings,
where small deviations from the subtle SM gauge cancellations can lead to
significant effects on physical observables -- these can be signals
of ``new physics''.  

Since the W's are unstable and short-lived particles, the W-pairs are not
observed directly in the experiments but through their decay products:
four-fermion (4f) final states (which may then also decay, radiate 
gluons/photons, hadronize, etc.). As high energy charged particles are
involved in the process, one can also observe energetic radiative photons.  
So, at the parton level, one has to consider a general process:
\vspace{-2mm}
\begin{equation}
e^+ + e^- \longrightarrow 4f + n\gamma, \:\: (n=0,1,2,\ldots),
\label{wp:4f-proc}
\end{equation}
where also some background (non-WW) processes contribute.
In a theoretical description of this process -- according to quantum field
theory -- one also has to include virtual effects, the so-called loop 
corrections. This general process is very complicated since it involves
$\sim 80$ different channels (4f final states) with complex peaking behaviour
in multiparticle phase space and a large number of Feynman diagrams to be
evaluated. Even in the massless-fermion approximation the number of Feynman
graphs grows up from $9$--$56$ per channel at the Born level to an enormous 
$3579$--$15948$ at the one-loop level~\cite{ww-LEP2YR:1996}. 
The full one-loop calculations have not been finished yet, even for the 
simplest case (doubly plus singly W-resonant diagrams)~\cite{vicini:1998}. 
But even if they
existed one would be faced with problems in their numerical evaluations 
in practical applications, particularly within Monte Carlo event generators
-- they would be extremely sizeable and very slow.
These are the reasons why efficient approximations in the theoretical 
description of this process are necessary. These approximations should
be such that on the one hand they would include all contributions/corrections
that are necessary for the required theoretical accuracy (dependent on 
experimental precision) and on the other hand they would be efficient enough 
for numerical computations. Given the complicated topologies of the 4f 
($+ n\gamma$) final states, such calculations should be, preferably, given 
in terms of a Monte Carlo event generator that would allow one to 
simulate the process directly~\cite{mc-LEP2YR:1996,4f-LEP2YR:2000}. 
Here we present
such a solution for the W-pair production process, which consists of two 
complementary Monte Carlo event generators: {\tt YFSWW3} and {\tt KoralW}.
More details on {\tt YFSWW3} can be found in 
Refs.~\cite{yfsww2:1996,yfsww3:1998,yfsww3:2000,yfsww3:2000a} and on 
{\tt KoralW} in Refs.~\cite{koralw:1996a,koralw:1996b,koralw:1999}.

{\tt KoralW} includes the full lowest-order $e^+e^- \rightarrow 4f$ process 
but with simplified radiative corrections -- the universal ones such as 
initial-state radiation (ISR), the Coulomb effect, etc. In {\tt YFSWW3}, 
on the other hand, the lowest-order process is simplified -- only the doubly 
W-resonant contributions are taken into account, but inclusion of the 
radiative corrections in this process goes beyond the universal ones.
In the current version of {\tt YFSWW3} only those non-universal (non-leading) 
corrections are included that are necessary to achieve the theoretical 
precision for the total WW cross section of $0.5\%$ required at LEP2. 
For the future linear colliders (LC) this may not be sufficient, so even 
some higher-order corrections would have to be added, which is possible 
within the framework of {\tt YFSWW3}. 
The important thing is that the two programs have a well
established common part, which is the doubly W-resonant (WW) process with
the same universal radiative corrections. This, as will be shown later,
allows us to combine the results of the two programs to achieve the desired
theoretical precision for WW observables. The ISR effects in both programs
are based on the Yennie--Frautschi--Suura (YFS) exclusive exponentiation 
procedure~\cite{yfs:1961,yfs2:1990}, with an arbitrary number of non-zero 
$p_T$ radiative photons. The Coulomb correction is implemented in the
standard version according to Ref.~\cite{coul:1995} and also in the form of 
the ``screened'' Coulomb ansatz of Ref.~\cite{scc:1999}, which is an
efficient approximation of non-factorizable corrections. 
The full 4f matrix element with non-zero fermion masses for {\tt KoralW} has 
been generated using the GRACE system of the MINAMI-TATEYA 
collaboration~\cite{grace:1994}. For an efficient event generation, two
independent 4f phase-space presamplers have been developed~\cite{MS-ZW:2000}.
In this way {\tt KoralW} is able to provide the important 4f-background
correction to the WW-process in the form of MC events. However, as was
already shown in Ref.~\cite{ww-LEP2YR:1996}, the pure universal radiative 
corrections and the 4f-background corrections are not sufficient for
a final theoretical precision tag of $0.5\%$ for LEP2 experiments.  
By using the exact ${\cal O}(\alpha)$ calculations of 
Refs.~\cite{bohm:ewc,FJKZ:ewc} for on-shell W-pair production, 
it was shown that the non-leading electroweak (EW) corrections
can be as large as $1$--$2\%$ at LEP2 energies (as will be seen later, they
are even larger at LC energies). These calculations were done, however,
in the on-shell-W approximation (stable W), so the question was how to 
implement (or extend) them in the realistic off-shell WW production.
A workable solution to this turned out to be the so-called leading-pole
approximation (LPA). The LPA was also needed for other reasons. Namely, the
matrix element for the WW production and decay based on three double-resonant
Feynman graphs (so-called CC03) is not SU(2)$_L\times$U$_1$ gauge-invariant,
and the simplest way to achieve the full gauge invariance is to use the LPA. 

There are two approaches within the LPA: the one already discussed in 
Ref.~\cite{ww-LEP2YR:1996}
and employed in the actual calculations for the WW process in 
Ref.~\cite{frits:1999}, and the second advocated by R.~Stuart in 
Ref.~\cite{stuart:1995}. In the first approach, the whole matrix element
is expanded in Laurent series about complex poles corresponding to two resonant
W's; then in the LPA only the leading terms of this expansion are retained.
In this approach one gets a direct correspondence to the on-shell W-pair
production and decay, but the results can differ from the realistic process
by several per cent. This can be corrected by adding the difference between
the predictions of the full 4f process and this approximation, at least at the 
Born level; however, it is not obvious how to do it on an event-by-event 
basis. We have implemented in {\tt YFSWW3} this solution 
and it is called the LPA$_b$ option -- it can
be useful for some tests/cross-checks. In the second approach, the 
gauge-invariant matrix element is first decomposed into a sum of Lorentz scalar
functions multiplied by spinor and Lorentz-tensor factors according to the
standard $S$-matrix theory~\cite{eolp:1966}. Then, only the Lorenz scalar
functions, which describe the finite-range W propagation, are expanded about
their complex poles. In the LPA, as previously done, only the 
leading terms in $(\Gamma_W/M_W)$ are 
retained. In this approach the results are very close to the predictions
based on the minimum gauge-invariant subset of Feynman diagrams including
the WW production (so-called CC11), e.g. for the total cross section
the differences are below $0.1\%$ at 200~GeV and $\sim 0.5\%$ at 500~GeV.
This solution is implemented in {\tt YFSWW3} as the LPA$_a$ option and
it is {\em recommended} for the event generation.  
The non-universal (non-leading) corrections are included in both LPAs through
the YFS exponentiation for the WW production stage including photon radiation
off the W bosons (split in a gauge-invariant way into the radiation in the
production and decay stages). Here we employ the exact ${\cal O}(\alpha)$ 
calculations for the on-shell WW production of Ref.~\cite{FJKZ:ewc}.
In the on-pole LPA residuals we make the approximation $s_p\approx M_W^2$,
where $s_p$ is the complex pole position and $M_W$ is the on-shell W mass,
which means neglecting terms $\sim(\alpha/\pi)(\Gamma_W/M_W)$ -- 
unimportant for the aimed theoretical accuracy. For the radiation in the
W decays, we use in the current version of {\tt YFSWW3} the leading-log-type 
program PHOTOS~\cite{photos:1994}, normalized to the radiatively corrected
W branching ratios; however, the YFS exponentiation for this process is in
progress. The non-factorizable corrections (interferences between various
stages of the process) have been included only via the so-called screened
Coulomb ansatz~\cite{scc:1999} (which is a sufficient approximation for
LEP2), but can be implemented to their full extent in the future.

Having these two MC event generators, we can combine their results, 
in order to obtain precise predictions for the WW process, in two ways. 
Either we can take the best prediction from {\tt YFSWW3} and correct it
for the 4f background using {\tt KoralW}, which can be symbolically
denoted by:
\vspace{-3mm}
\begin{equation}
\sigma_{Y/K} = \: \sigma_Y \: \oplus \: \delta_{K}^{4f}\;,
\label{wp:syk}
\end{equation}
or we can take the best prediction from {\tt KoralW} and correct it
for the non-leading (NL) effects to the ``signal'' process from  {\tt YFSWW3},
which we can write symbolically as:
\vspace{-5mm}
\begin{equation}
\sigma_{K/Y} = \: \sigma_K \: \oplus \: \delta_{Y}^{NL}\;.
\label{wp:sky}
\end{equation}
This can be done easily at the level of the total cross section as well as
for the differential distributions. Recently, reweighting interfaces have been
developed for the two programs so that it can also be done on an 
event-by-event basis~\cite{yfsww3-1.15,koralw-1.42.3}. 
All this is possible because both programs have some
common basic distribution, which is the WW signal process with the universal
radiative correction, and it has been checked that they agree
very well at this level~\cite{yfsww3:2000a}.

{\tt YFSWW3} was also compared with an independent MC program, 
{\sc RacoonWW} \cite{racww:2000}, which
includes the non-universal ${\cal O}(\alpha)$ corrections for
the W-pair production. The two programs were found to agree for the total
WW cross section $< 0.4\%$ at LEP2 energies~\cite{4f-LEP2YR:2000} 
and $< 0.5\%$ at 500~GeV~\cite{yfsww3:2000a}. 
Numerically, the non-universal ${\cal O}(\alpha)$ corrections as calculated
by {\tt YFSWW3} are $\sim 1$--$2\%$ at LEP2 energies and $\sim 5$--$10\%$
at LC energies ($0.5$--$1.5$~TeV), and they are always negative. On the other
hand the ISR corrections change their sign from being large negative near
the WW threshold to being large positive at LC energies (thus cancelling
partially the effects of non-universal corrections).

\begin{center}
{\large\bf  Acknowledgements}
\end{center}

\noindent
One of us (W.P.) thanks the organizers of the LCWS~2000 at Fermilab
for their kind hospitality and their financial support. 
We also acknowledge the support of the CERN Theory Division,
all the LEP Collaborations and the DESY Directorate.


\vspace{-2mm}
\begin{references}
%
\bibitem{LEP2YR:1996}
{\em Physics at LEP2},
edited by G. Altarelli, T. Sj\"ostrand and F. Zwirner 
(CERN 96-01, Geneva, 1996), 2 vols.
%
\bibitem{ww-LEP2YR:1996}
W. Beenakker {\it et al.},
{\em WW Cross-Sections and Distributions}, 
in~\cite{LEP2YR:1996}, Vol.~1, p.~79.
%
\bibitem{vicini:1998}
A. Vicini, Acta Phys. Polon., {\bf B29} (1998) 2847.
%
\bibitem{mc-LEP2YR:1996}
D. Bardin {\it et al.},
{\em WW Event Generators for WW Physics}, 
in~\cite{LEP2YR:1996}, Vol.~2, p.~3.
%
\bibitem{4f-LEP2YR:2000} 
M. Gr\"unewald {\it et al.},
{\em Four-Fermion Production in Electron-Positron Collisions}, 
in: {\it Reports of the Working Groups on Precision Calculations for
LEP2 Physics}, edited by S.~Jadach, G.~Passarino and R.~Pittau
(CERN 2000-009, Geneva, 2000), p.~1.
%
\bibitem{yfsww2:1996}
S. Jadach, W. P{\l}aczek, M. Skrzypek and B.F.L. Ward,
Phys. Rev. {\bf D54} (1996) 5434.
%
\bibitem{yfsww3:1998}
S. Jadach, W. P{\l}aczek, M. Skrzypek, B.F.L. Ward and Z. W\c{a}s,
Phys. Lett. {\bf B417} (1998) 326. 
%
\bibitem{yfsww3:2000}
S. Jadach, W. P{\l}aczek, M. Skrzypek, B.F.L. Ward and Z. W\c{a}s,
Phys. Rev. {\bf D61} (2000) 113010;
preprint CERN-TH/99-222; hep-ph/9907436.
%
\bibitem{yfsww3:2000a}
S. Jadach, W. P{\l}aczek, M. Skrzypek, B.F.L. Ward and Z. W\c{a}s,
preprint CERN-TH/2000-337; hep-ph/0007012;
to be submitted to Phys. Lett. {\bf B}.
%
\bibitem{koralw:1996a}
M. Skrzypek, S. Jadach, W. P{\l}aczek and Z. W\c{a}s,
Comput. Phys. Commun. {\bf 94} (1996) 216.
%
\bibitem{koralw:1996b}
M. Skrzypek {\it et al.},
Phys. Lett. {\bf B372} (1996) 289. 
%
\bibitem{koralw:1999}
S. Jadach, W. P{\l}aczek, M. Skrzypek, B.F.L. Ward and Z. W\c{a}s,
Comput. Phys. Commun. {\bf 119} (1999) 272.
%
\bibitem{yfs:1961}
D.R. Yennie, S. Frautschi and H. Suura, Ann. Phys. (NY) {\bf 13} (1961) 379.
%
\bibitem{yfs2:1990}
S. Jadach and B.F.L. Ward,
Comput. Phys. Commun. {\bf 56} (1990) 351.
%
\bibitem{coul:1995}
V.S. Fadin, V.A. Khoze, A.D. Martin and W.J. Stirling,
Phys. Lett. {\bf B363} (1995) 112.
%
\bibitem{scc:1999}
A. P. Chapovsky and V. A. Khoze, Eur. Phys. J. {\bf C9} (1999) 449; 
hep-ph/9902343.
%
\bibitem{grace:1994}
J. Fujimoto {\it et al.}, 
{\em GRACE User's manual, version 2.0},
MINAMI-TATEYA collaboration, 1994.
%
\bibitem{MS-ZW:2000}
M. Skrzypek and Z. W\c{a}s, 
Comput. Phys. Commun. {\bf 125} (2000) 8. 
%
\bibitem{bohm:ewc}
M. B\"ohm {\it et al.}, Nucl. Phys. {\bf B304} (1988) 463.
%
\bibitem{FJKZ:ewc}
J. Fleischer,  F. Jegerlehner and M. Zra\l{}ek, 
Z. Phys. {\bf C42} (1989) 409;\\
M. Zra\l{}ek and K. Ko\l{}odziej, Phys. Rev. {\bf D43} (1991) 43; \\
J. Fleischer, K. Ko\l{}odziej and F. Jegerlehner, 
Phys. Rev. {\bf D47} (1993) 830; \\
J. Fleischer {\it et al.}, Comput. Phys. Commun. {\bf 85} (1995) 29.
%
\bibitem{frits:1999}
W. Beenakker, F.A. Berends and A.P. Chapovsky, 
Nucl. Phys. {\bf B548} (1999) 3. 
%
\bibitem{stuart:1995}
R.G. Stuart, Nucl. Phys. {\bf B498} (1997) 28; 
Eur. Phys. J. {\bf C4} (1998) 259; hep-ph/9706431, 9706550.
%
\bibitem{eolp:1966}
R. J. Eden, P.V. Landshoff, D.I. Olive, and J.C. Polkinghorne,
{\em The Analytic S-Matrix}
(Cambridge University Press, Cambridge, 1966).
%
\bibitem{photos:1994}
E. Barberio and Z. W\c{a}s, Comput. Phys. Commun. {\bf 79} (1994) 291.
%
\bibitem{yfsww3-1.15}
S. Jadach, W. P{\l}aczek, M. Skrzypek, B.F.L. Ward and Z. W\c{a}s,
{\tt YFSWW3 version 1.15}, available at:
{\tt http://cern.ch/placzek/}.
%
\bibitem{koralw-1.42.3}
S. Jadach, W. P{\l}aczek, M. Skrzypek, B.F.L. Ward and Z. W\c{a}s,
{\tt KoralW}, available at: 
{\tt http://hpjmiady.ifj.edu.pl/}.
%
\bibitem{racww:2000}
A. Denner, S. Dittmaier, M. Roth and  D. Wackeroth,
Nucl. Phys. {\bf B587} (2000) 67. 
\end{references}
\end{document}